\documentclass[eqsecnum,twocolumn,showpacs,aps]{revtex4}

\usepackage{graphicx}
\usepackage{rotating}

\def\lapprox{{\raise0.5ex\hbox{$<$}\hskip-0.80em\lower0.5ex\hbox{$\sim$}
}}
\def\gapprox{{\raise0.5ex\hbox{$>$}\hskip-0.80em\lower0.5ex\hbox{$\sim$}
}}

\begin{document}
\title{Double-Pionic Fusion of Nuclear Systems and the ABC
  Effect ---\\ Approaching a Puzzle by
  Kinematically  Complete Measurements}  
\author{M. Bashkanov$^1$, C. Bargholtz$^{11}$, 
M. Ber{\l}owski$^6$,
D. Bogoslawsky$^2$,
H. Cal\'en$^3$, 
 H. Clement$^1$, L.~Demiroers$^5$, 
E. Doroshkevich$^1$, D.~Duniec$^4$,
C. Ekstr\"om$^3$,
K. Fransson$^3$, 
L. Geren$^{11}$
L. Gustafsson$^4$,
B. H\"oistad$^4$,
G. Ivanov$^2$,
M. Jacewicz$^4$,
E. Jiganov$^2$,
T. Johansson$^4$, 
O. Khakimova$^1$,
S. Keleta$^4$,
I. Koch$^4$,
F. Kren$^1$, 
S. Kullander$^4$,
A. Kup\'s\'c$^3$,
K. Lindberg$^{11}$, P. Marciniewski$^3$,
R. Meier$^1$,
B. Morosov$^2$,
C. Pauly$^7$, H.~Pettersson$^4$,
Y. Petukhov$^2$,
A. Povtorejko$^2$, A.~Pricking$^1$,
R.J.M.Y. Ruber$^3$, 
K. Sch\"onning$^4$,
W. Scobel$^5$,
B. Shwartz$^{8}$,
T. Skorodko$^1$, 
V. Sopov$^{10}$,
J. Stepaniak$^6$,
P.-E.~Tegner$^{11}$,
P. Th\"orngren-Engblom$^4$,
V. Tikhomirov$^2$,
A. Turowiecki$^{9}$,
G.J. Wagner$^1$, 
M. Wolke$^7$,
J. Zabierowski$^{7}$,
I. Zartova$^{11}$,
J. Z{\l}omanczuk$^4$}
\affiliation{
$^1$~Physikalisches Institut der Universit\"at T\"ubingen, D-72076
  T\"ubingen, Germany \\
$^2$~Joint Institute for Nuclear Research, Dubna, Russia \\
$^3$~The Svedberg Laboratory, Uppsala, Sweden \\
$^4$~Uppsala University, Uppsala,Sweden \\
$^5$~Hamburg University, Hamburg, Germany \\
$^6$~Soltan Institute of Nuclear Studies, Warsaw and Lodz, Poland \\
$^7$~Forschungszentrum J\"ulich, Germany \\
$^{8}~$Budker Institute of Nuclear Physics, Novosibirsk, Russia \\
$^{9}$~Institute of Experimental Physics, Warsaw, Poland \\
$^{10}$~Institute of Theoretical and Experimental Physics, Moscow,
  Russia \\ 
$^{11}$~Department of Physics, Stockholm University, Stockholm, Sweden \\
~~~~~~~~ \\
(CELSIUS/WASA Collaboration)}

\date{\today}

\begin{abstract}
The ABC effect --- 
a puzzling low-mass enhancement in the $\pi\pi$ invariant mass spectrum --- is
well-known from inclusive measurements of two-pion production in nuclear
fusion reactions. Here we report on first exclusive and kinematically complete
measurements of the most basic double pionic fusion reaction $pn \to
d\pi^0\pi^0$ at beam energies of 1.03 and 1.35 GeV. The measurements, which
have been carried out at CELSIUS-WASA, reveal the ABC effect  
to be a $(\pi\pi)_{I=L=0}$ channel phenomenon associated with  both a
resonance-like energy dependence in the integral cross section and the 
formation of a $\Delta\Delta$ system in the intermediate state. A
corresponding simple {\it s}-channel resonance ansatz provides a surprisingly
good description of the data.
\end{abstract}


\maketitle
The ABC effect --- first observed by Abashian, Booth
and Crowe \cite{abc} in
the double pionic fusion of deuterons and protons to $^3$He --- stands for an
unexpected enhancement at low masses in the invariant $\pi\pi$ mass spectrum
$M_{\pi\pi}$. Follow-up experiments
\cite{ban,hom,hal,bar,plo,abd,col,ban1,cod,wur} revealed this
effect to be of isoscalar nature and to show up in cases, when the two-pion
production process leads to a bound nuclear system. With
the exception of low-statistics bubble-chamber measurements \cite{bar,abd} all 
experiments conducted on this issue have been inclusive measurements carried
out preferentially with single-arm magnetic spectrographs for the detection
of the fused nuclei.

Initially the low-mass enhancement had been interpreted as an unusually large 
$\pi\pi$ scattering length and evidence for the $\sigma$ meson, respectively
\cite{abc}. Since the effect showed up particularily clear at beam energies
corresponding to the excitation of two $\Delta$s in the nuclear system, the ABC
effect was interpreted later on by a $t$-channel $\Delta\Delta$
excitation in the course of the reaction process leading to both a low-mass
and a high-mass enhancement in isoscalar $M_{\pi\pi}$ spectra
\cite{ris,barn,anj,gar,mos,alv}. In fact, the
missing momentum spectra from inclusive measurements have been in support of
such predictions. 


In order to shed more light on this issue and complementing our previous
result on the double-pionic fusion to $^3$He \cite{bash,MB} exclusive and
kinematically overconstrained measurements have been carried out on the most
basic system for double-pionic fusion, the $pn
\rightarrow d\pi^0\pi^0$ reaction. To this end we have measured this reaction
in the quasi-free mode $pd \to p_{spectator}d\pi^0\pi^0$ at beam energies
$T_p$ = 1.03 and 1.35 GeV at CELSIUS using 
the 4$\pi$ WASA detector setup including the pellet target system \cite{barg}.
The latter provides frozen deuterium pellets of size $\approx 20 \mu m$, which
cross the beam perpendicularly with a frequency of $\approx 7 kHz$.
The beam energies have been chosen to be in the region of the ABC effect
as known from inclusive measurements.
The experimental results on the $pd \rightarrow
^3$He$\pi^0\pi^0$ and $pd \rightarrow ^3$He$\pi^+\pi^-$ reactions and first
results from the measurements 
of the $dd \to ^4$He$\pi\pi$ reaction are given in Refs. \cite{bash, MB,sam}.

The deuterons emerging from the reaction of interest here have been detected
in the forward 
detector and identified by the $\Delta$E-E technique using corresponding
informations from quirl and range hodoscope, respectively (see, e.g. Fig.1 of
ref. \cite{bash}). Gammas from the $\pi^0$ decay have been
detected in the central detector.

This way the full four-momenta have been measured for all
particles of an event --- except for the very low-energetic spectator proton,
which did not reach any active detector element. Thus kinematic fits with 3
overconstraints could be performed for each event. 

Due to Fermi motion of the nucleons in the target deuteron 
the quasifree reaction proceeds via a range of effective collision
energies. Based on the measured energies in the exit channel and the thus
event-by-event reconstructed total energies in the pn system the data have been
binned into small ranges of effective beam energy, in order to reduce 
the kinematical smearing and also to allow the extraction of the energy
dependence in the total cross section.

The absolute normalisation of these data has been obtained by
relative normalisation to the quasifree single-pion production   
$pn \rightarrow d\pi^0$ \cite{bys} measured simultaneously with the same
particle trigger.
For the higher beam energy we used in addition the $pn \to d\eta$ production
process for an alternative calibration. Both
calibration methods agree within 10 $\%$. However, since at the higher energy
the phase space coverage is less complete  
than at the lower energy, the determination of the absolute cross section
depends somewhat on the model used in the MC simulation for acceptance and
efficiency correction. Hence the cross sections derived for the
higher beam energy are plotted in Fig. 4 with an increased uncertainty. Also
in order to avoid contaminations due to the increased $\pi\pi\pi$
production and other background we introduced in the analysis of the
high energy data kinematic constraints on the spectator proton before kinematic
fits. 

\begin{figure}
\begin{center}
\includegraphics[width=0.23\textwidth]{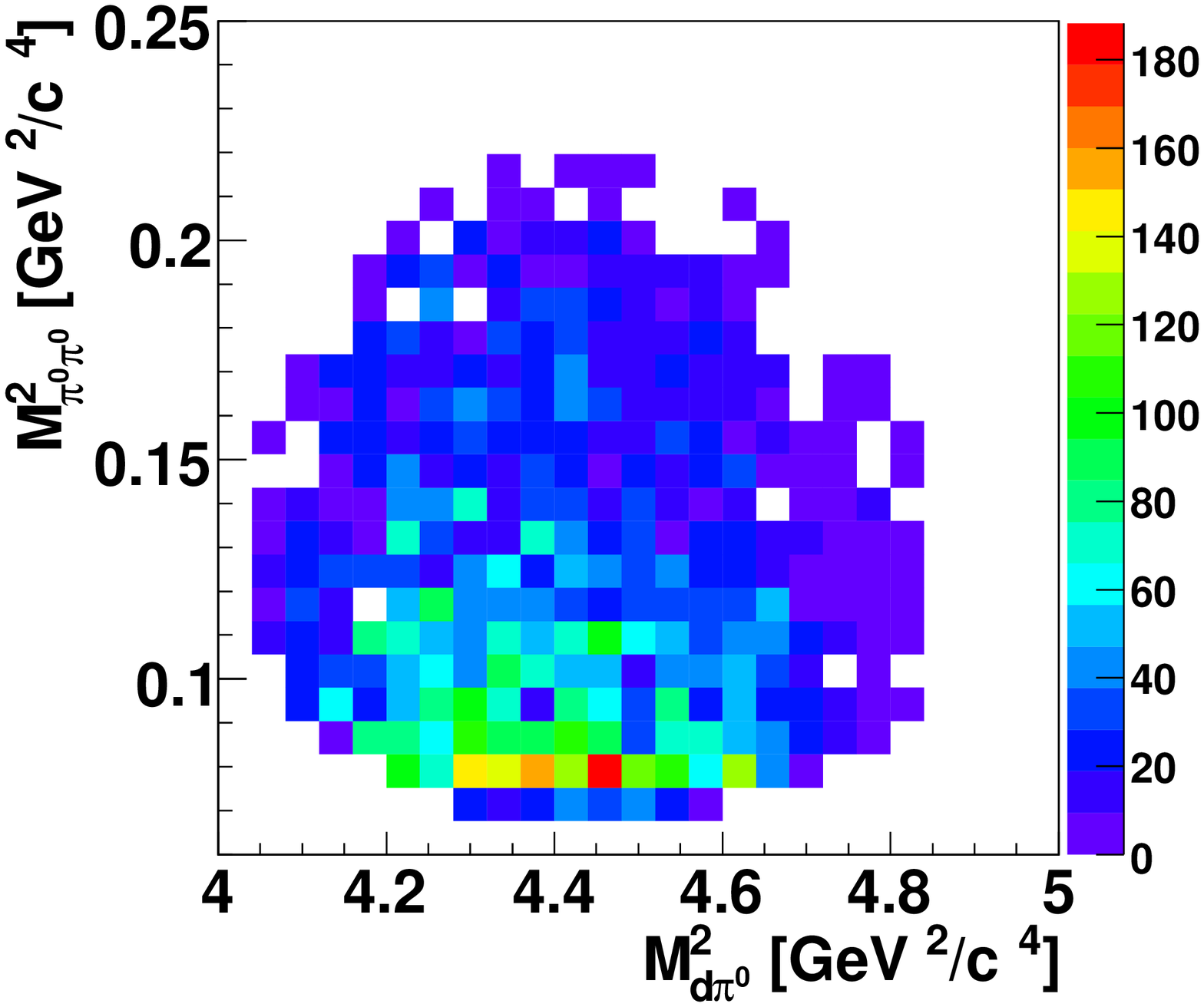}
\includegraphics[width=0.23\textwidth]{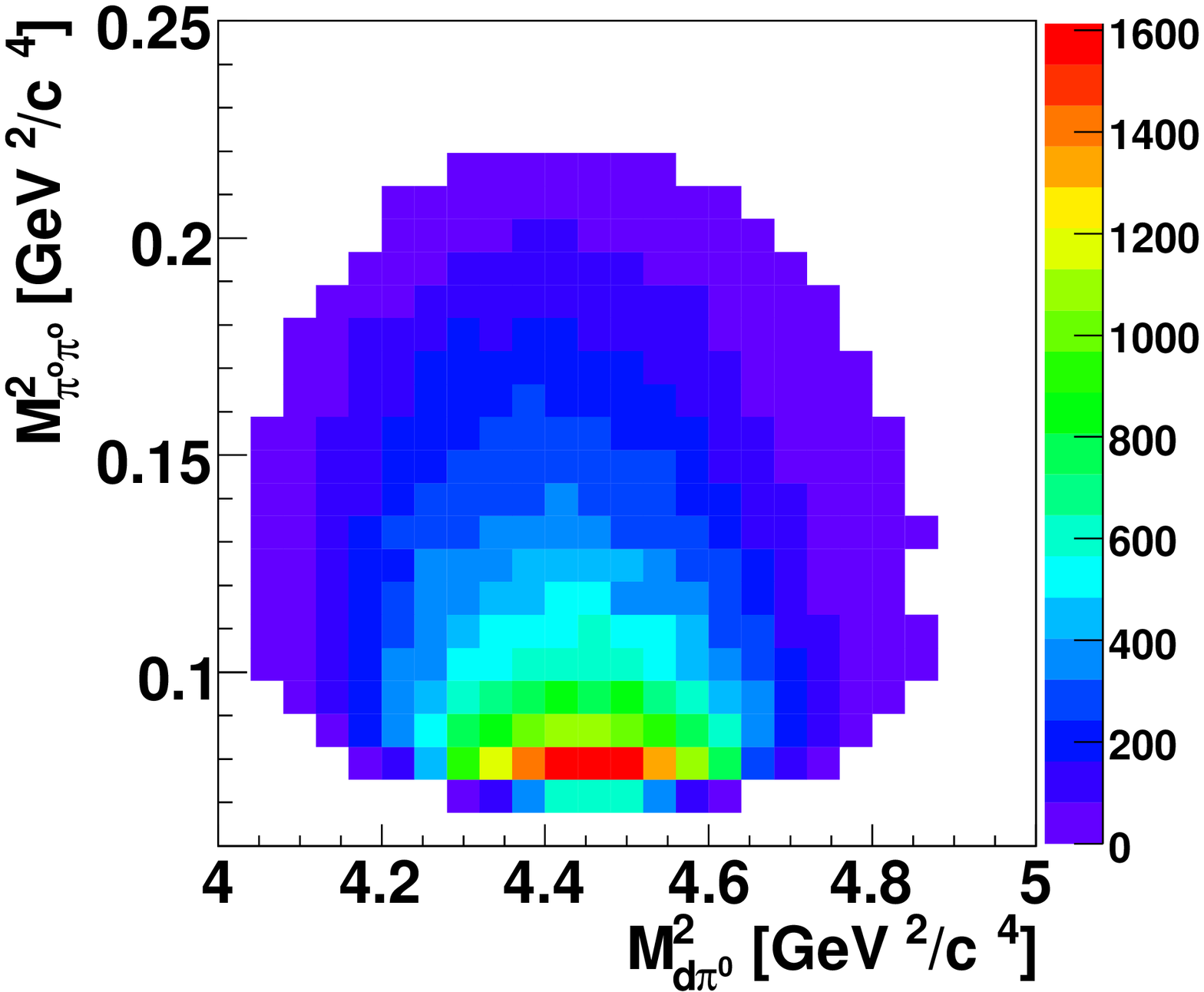}
\includegraphics[width=0.23\textwidth]{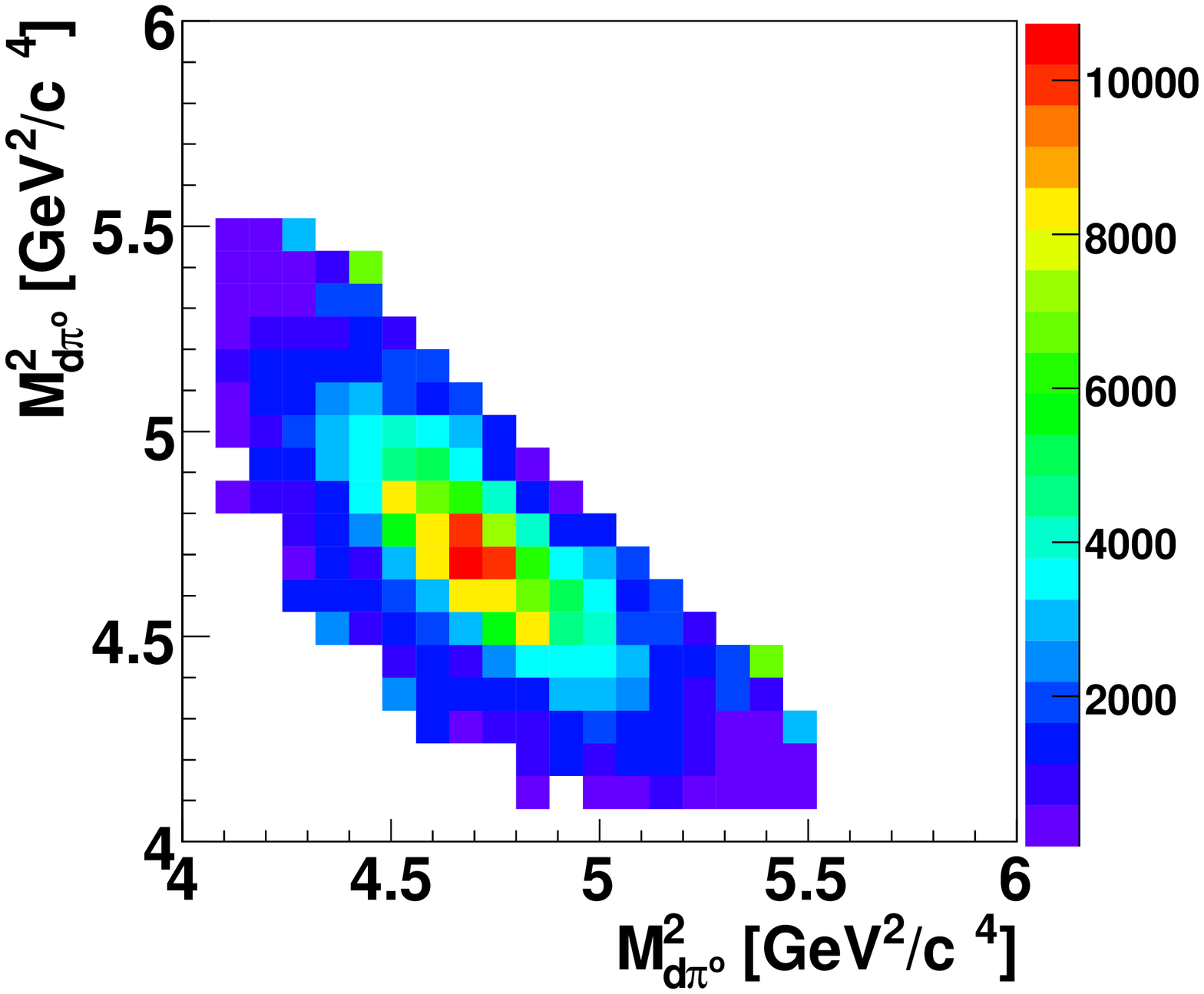}
\includegraphics[width=0.23\textwidth]{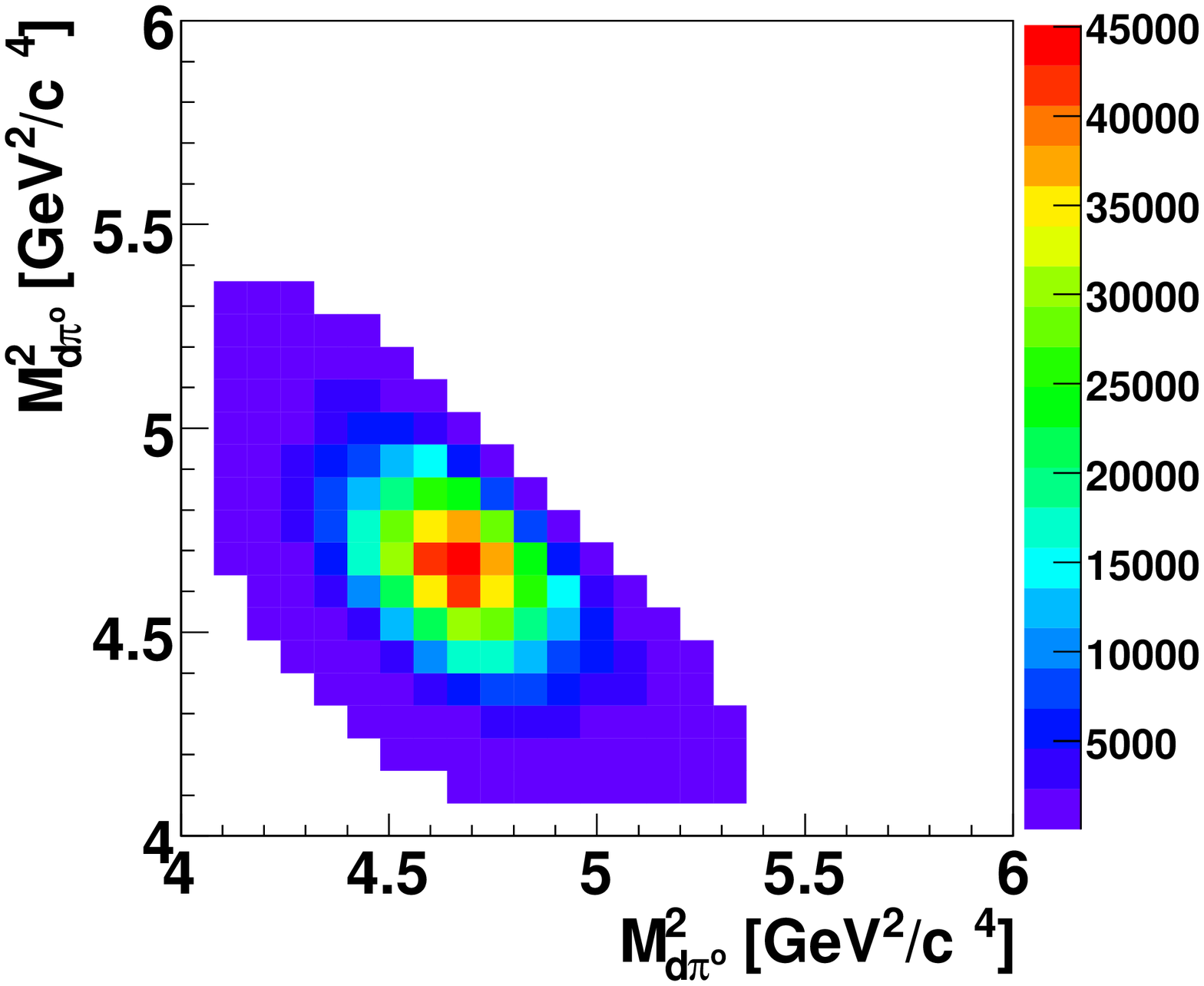}
\caption{Dalitz plot of the invariant mass distributions for 
$M_{d\pi^0}^2$ versus $M_{\pi^0\pi^0}^2$ (top) and $M_{d\pi^0}^2$ versus
$M_{d\pi^0}^2$ (bottom) for the quasifree reaction process
$pn \to 
  d\pi^0\pi^0$ in the range 1.00 - 1.03 GeV (top) and 1.27 - 1-37 GeV (bottom)
  of effective collision
  energies. Left: data, right: MC simulation of the reaction model, see text.
The color-coded scale (z-axis) is in arbitrary units}
\label{fig1}
\end{center}
\end{figure}


\begin{figure}[t]
\begin{center}
\includegraphics[width=0.23\textwidth]
                              {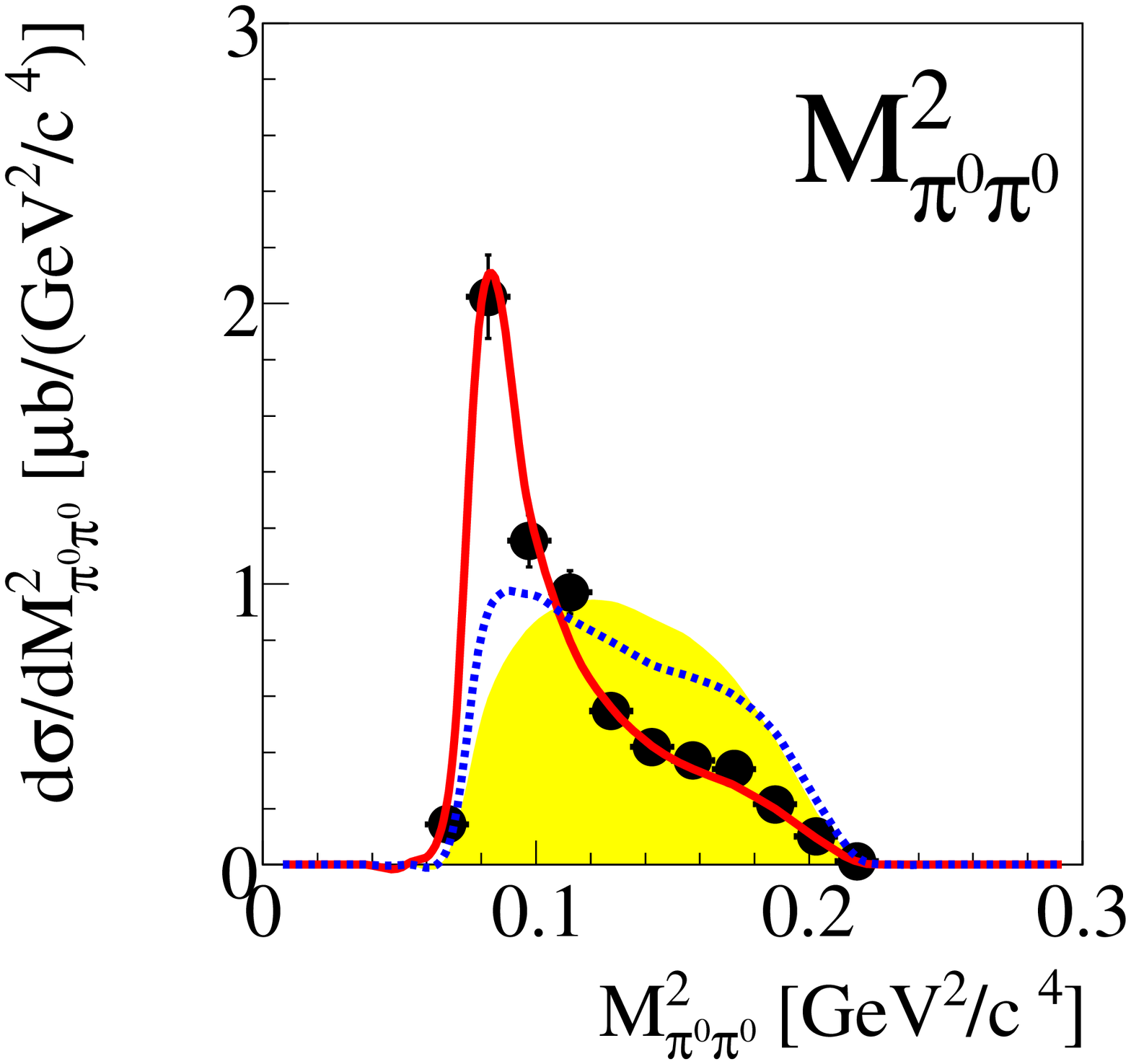}
\includegraphics[width=0.23\textwidth]
                               {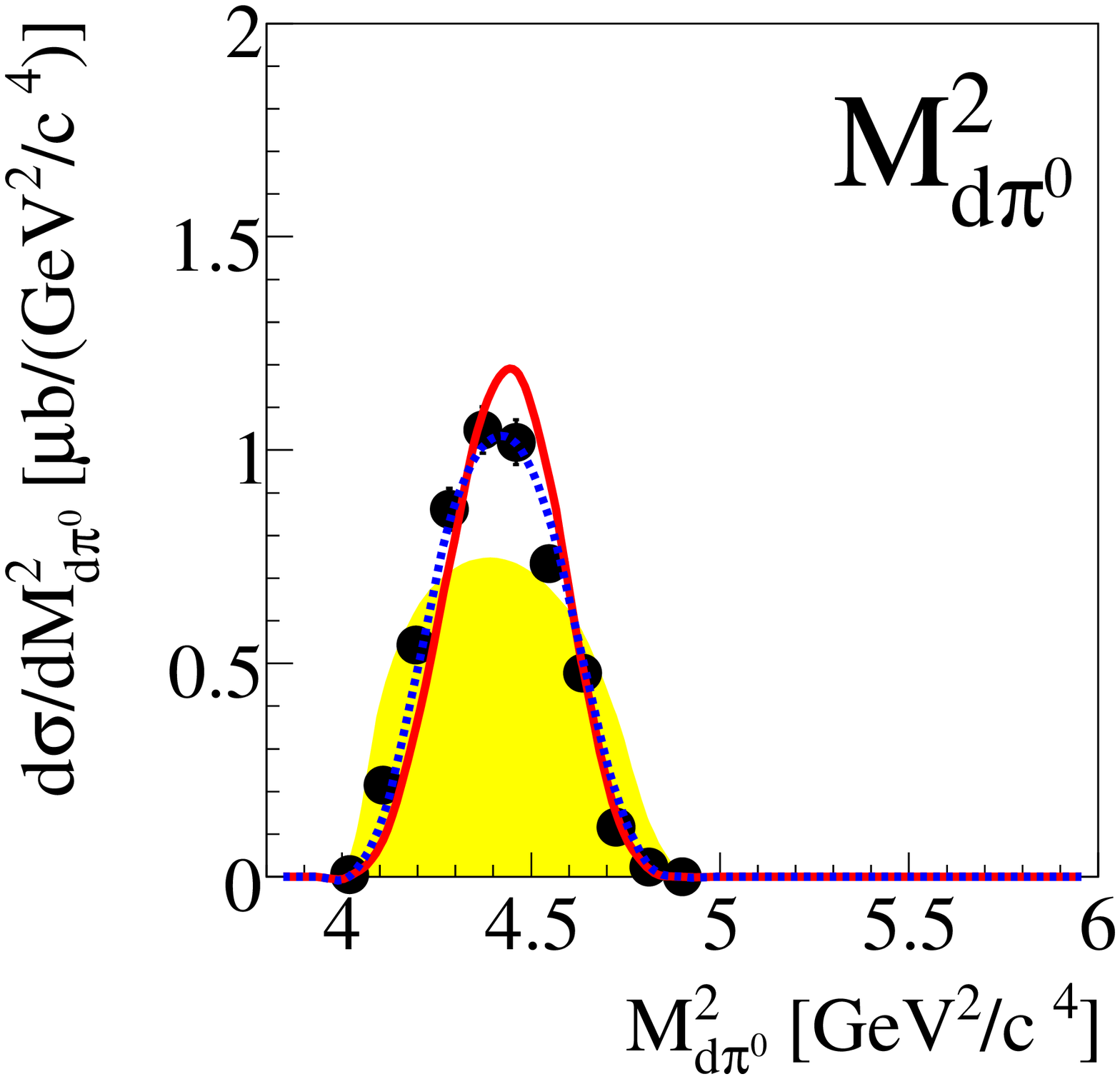}
\includegraphics[width=0.23\textwidth]
                          {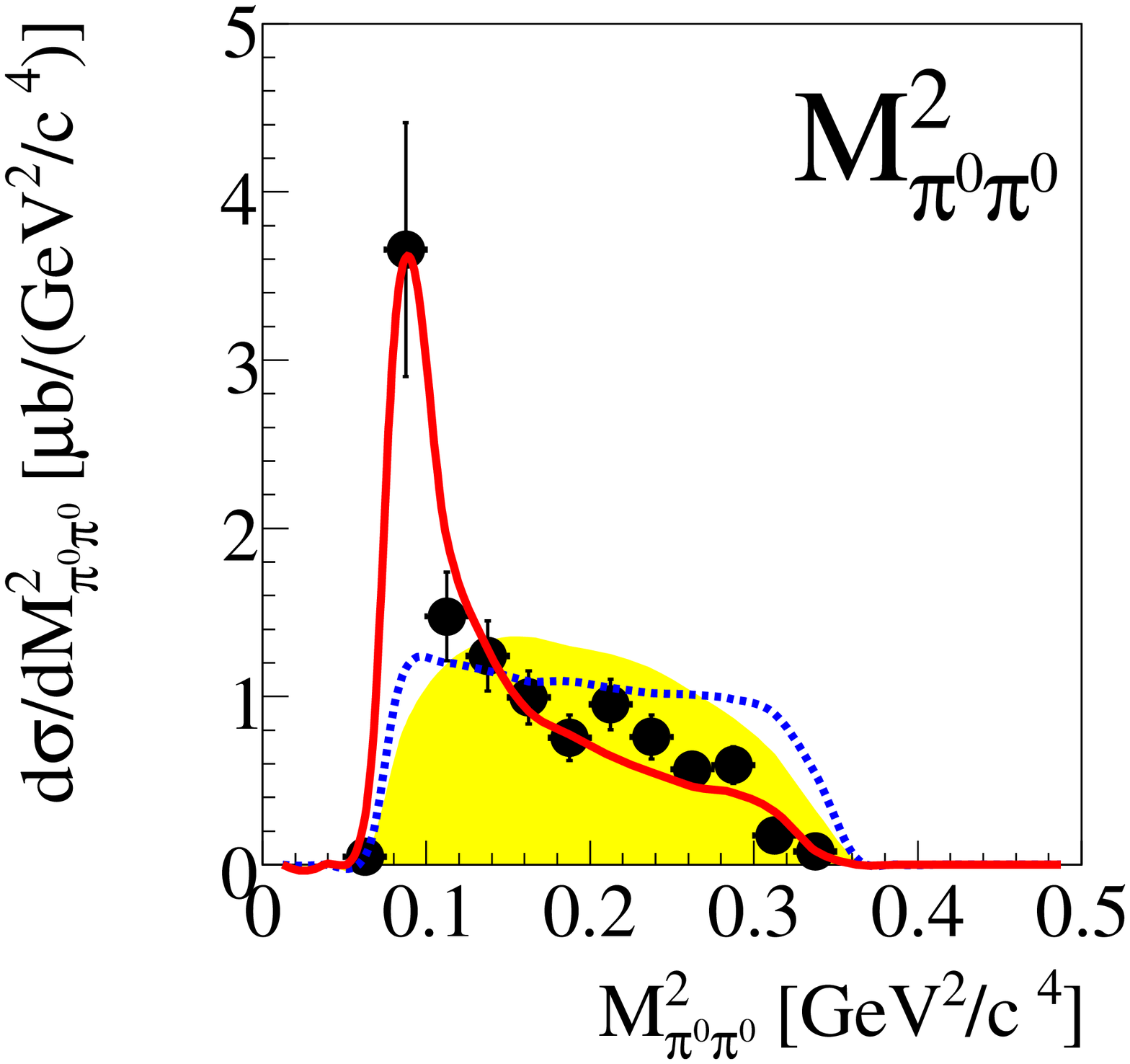}
\includegraphics[width=0.23\textwidth]
                           {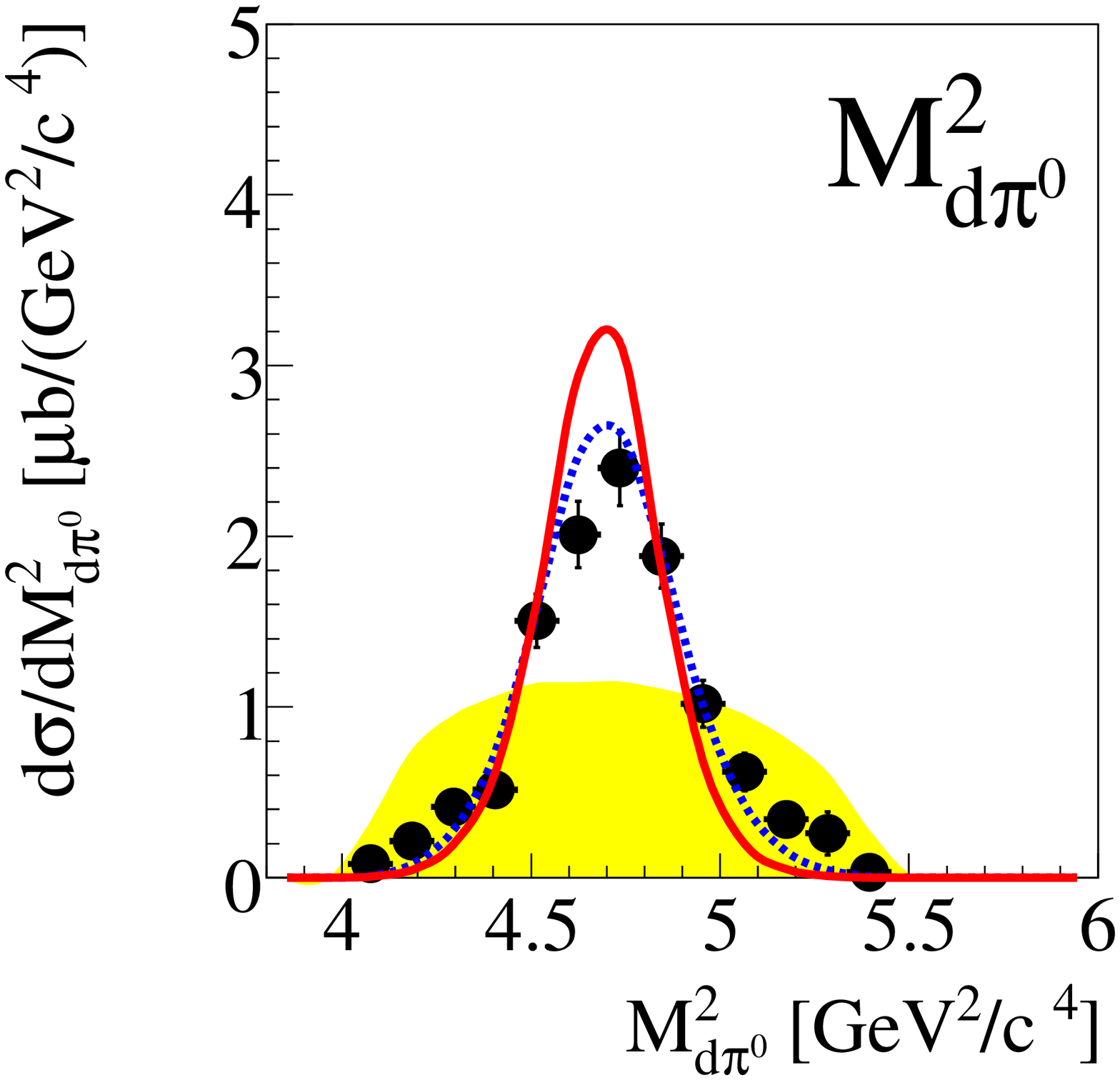}

\caption{
  Distributions of the invariant masses $M_{\pi^0\pi^0}$ and 
  $M_{d\pi^0}$ from the exclusive measurements (full dots)
  of the $pn \to d \pi^0\pi^0$ reaction at effective collision energies of
  1.00 - 1.03 (top) and 1.34 - 1.37 GeV (bottom).
  The shaded areas show the pure phase
  space distributions. Solid and dotted curves give
  $\Delta\Delta$ calculations with and without the assumption of a s-channel
  resonance. All curves are normalized to the experimental integral cross
  section. 
}
\label{fig3}
\end{center}
\end{figure}

Results of our measurements are shown in Figs. 1 - 4. 
Since the two $\pi^0$ particles emerging from the reaction of interest are
identical particles, observables depending only on a single $\pi^0$ 
like $M_{d\pi^0}$  are calculated by averaging over both possible combinations.
Figure 1 displays the
Dalitz plots of the invariant mass-squares $M_{d\pi^0}^2$ versus
$M_{\pi^0\pi^0}^2$ and of the invariant mass-square $M_{d\pi^0}^2$ of the
deuteron with one of the pions versus the same quantity with the other pion 
 --- 
both for data and Monte Carlo (MC) simulations of a model ansatz discussed
below. The Dalitz plots are far from being flat, i.e. far from being
phase-space like. They 
rather exhibit  very pronounced enhancements in the region of the $\Delta$
resonance and at the low-mass kinematic limit
of $M_{\pi^0\pi^0}^2$. This unusual phenomenon is known as the ABC effect. 
In this reaction the $\pi^0\pi^0$ channel is free
of both isovector ($I = 1$) and isotensor ($I = 2$) contributions due to
vanishing isospin coupling coefficients. Hence the
observation of the ABC effect here means that it must be of isoscalar ($I = 0$)
character.  Figure 2 depicts the Dalitz-plot projections, the spectra of
invariant mass-squares $M_{\pi^0\pi^0}^2$  and
$M_{d\pi^0}^2$  for the two bins of effective collision energies $T_p$ = 1.00 -
1.03 GeV and $T_p$ = 1.34 - 1.37 GeV. At both energies low-mass
$M_{\pi^0\pi^0}$ enhancement and $\Delta$ excitation, respectively, are the
predominating structures.  


\begin{figure} [h]
\begin{center}
\includegraphics[width=0.23\textwidth]{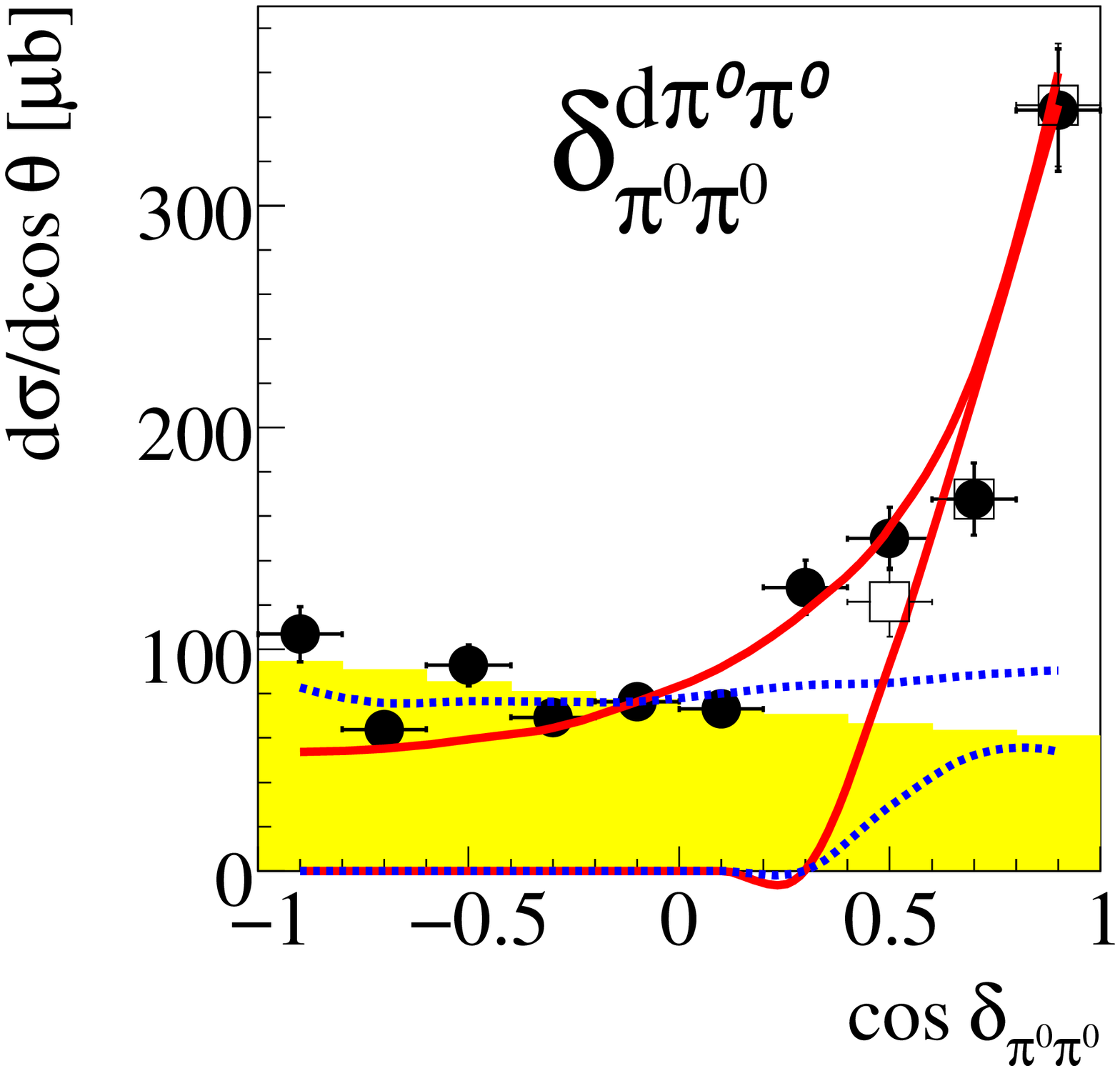}
\includegraphics[width=0.23\textwidth]{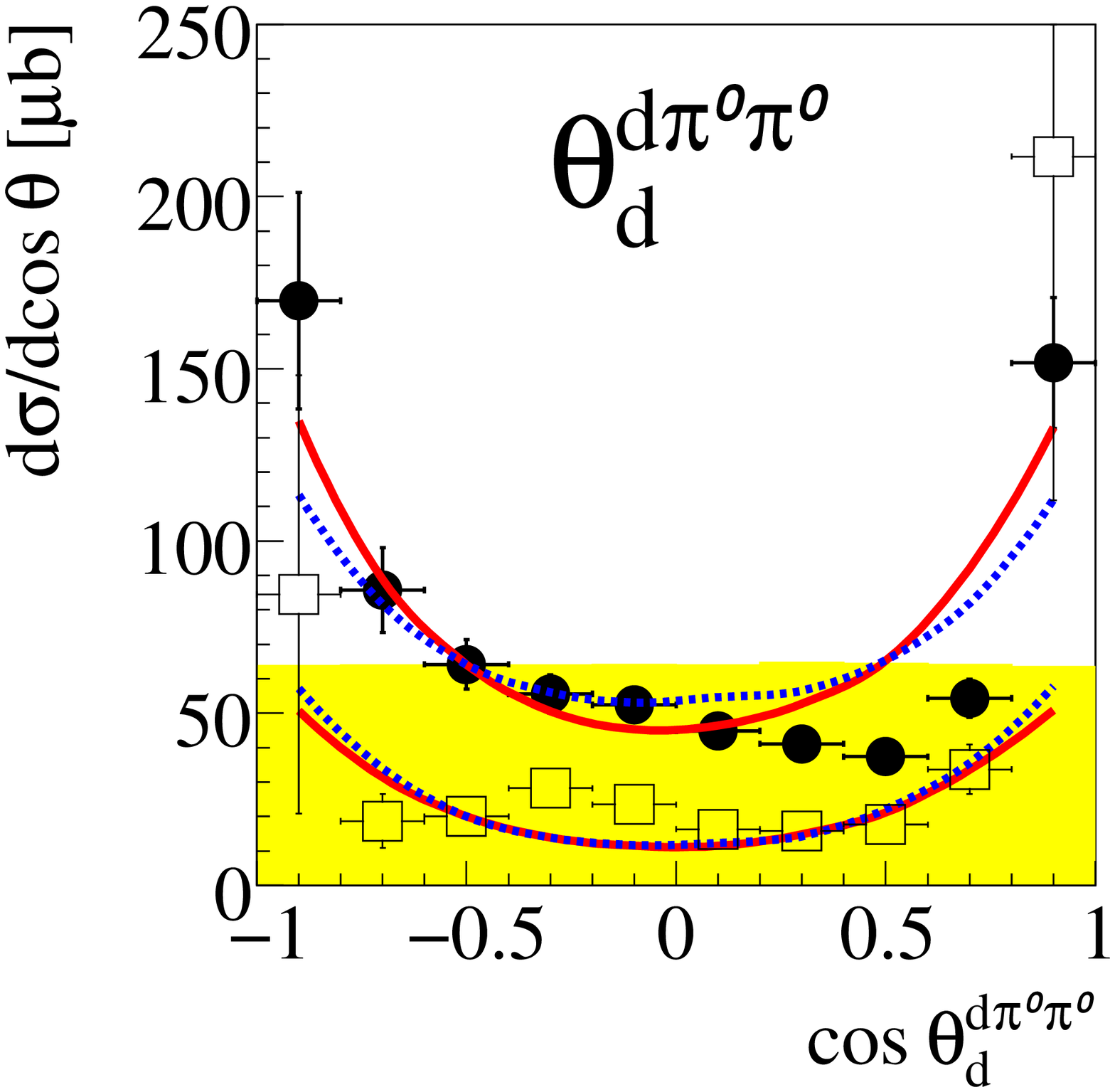}
\includegraphics[width=0.23\textwidth]{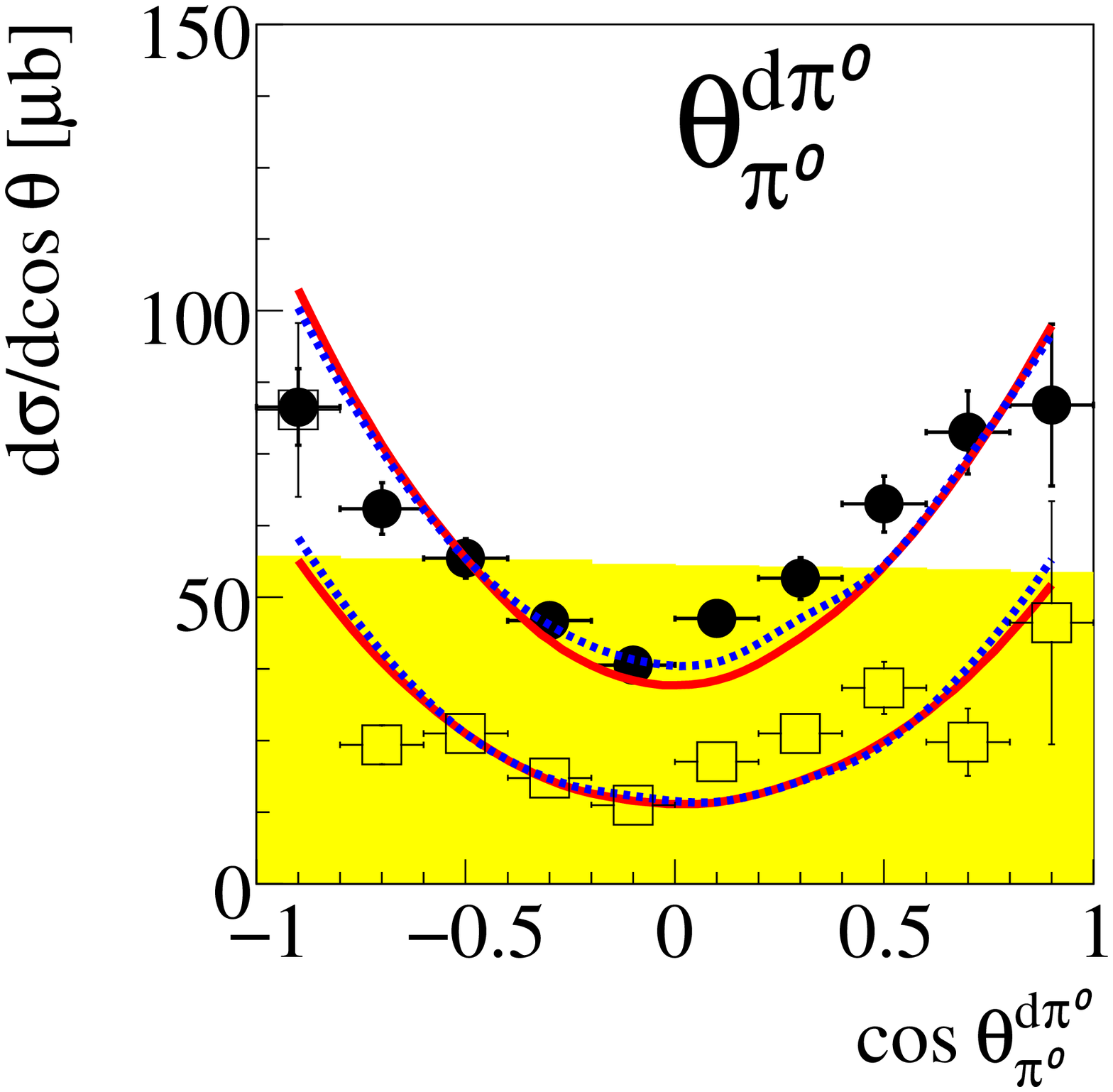}
\includegraphics[width=0.23\textwidth]{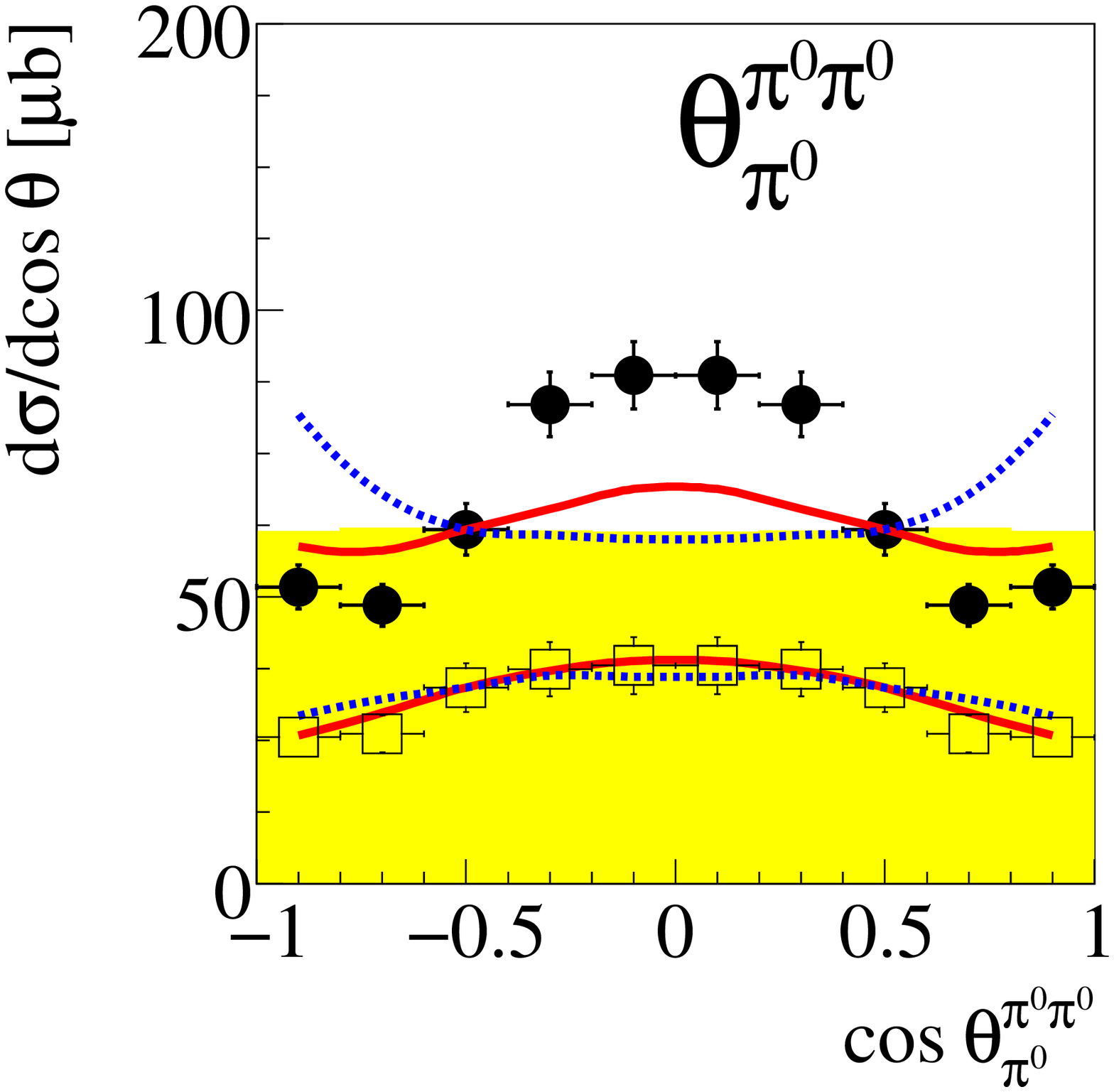}

\caption{Angular distributions in the reaction $pn \to
  d\pi^0\pi^0$ at $T_p$ =1.03 GeV for  
 the opening angle $\delta^{d\pi^0\pi0}_{\pi^0\pi^0}$ between the two pions, the angle of the
 deuteron $\Theta^{d\pi^0\pi0}_d$ - all in
 the $d\pi^0\pi^0$ cms - as well as the pion angles $\Theta_{\pi^0}^{d\pi^0}$ and
 $\Theta^{\pi\pi}_{\pi}$ in  $d\pi^0$ and   
 $\pi\pi$ subsystems (Jackson frame), respectively. The meaning of symbols and
 curves is as in Fig. 2. In addition data (open squares) and
 calculations are plotted also
 with the  constraint $M_{\pi\pi} < 0.32$ GeV/c$^2$.
}
\end{center}
\end{figure}

Angular distributions are shown in Fig. 3. The 
distribution of the opening angle $\delta_{\pi^0\pi^0}^{d\pi^0\pi^0}$ between
the two pions in the $d\pi^0\pi^0$ (or equivalently $pn$) center-of-mass
system (cms) peaks at small angles, in particular if we select events with
$M_{\pi^0\pi^0} \leq$~0.32 GeV/c$^2$. This means
that the low-mass enhancement is associated with pions leaving the interaction
vertex in parallel.
The distributions of the deuteron polar angles $\Theta_d^{d\pi^0\pi^0}$ in the
$d\pi^0\pi^0$ cms and of the pion polar angles $\Theta_{\pi^0}^{\pi^0\pi^0}$ and 
$\Theta_{\pi^0}^{d\pi^0}$ in $\pi^0\pi^0$ and $d\pi^0$ subsystems, 
respectively, are anisotropic and essentially symmetric about 
90$^{\circ}$. The anisotropy observed for the latter
corresponds just to the one expected from $\Delta$ decay. 
The anisotropy in the $\pi^0$ angular distribution in the $\pi\pi$ subsystem
signals some d-wave admixture. It  
vanishes, if we consider  only data with $M_{\pi^0\pi^0} \leq$ 0.32 
GeV/c$^2$, i.e. in the region of the low-mass enhancement. From this we  
deduce that the enhancement is of scalar nature, i.e. in total of 
scalar-isoscalar nature.


The observed low-mass enhancement is 
consistent with the findings in previous inclusive single-arm measurements,
where only the momentum of outgoing fused nuclei was measured. Note that 
the momentum of the fused ejectile is directly
related to the invariant $\pi\pi$ mass in the restricted kinematical range of
the measured scattering angle 
(see Figs. 3 - 5 in Ref. \cite{bash} and
Refs. \cite{abc,ban,hom,hal,bar,plo,ban1}).


The momentum spectra of inclusive single-arm measurements \cite{plo,col} 
exhibit also a strong central maximum (corresponding to large 
$M_{\pi\pi}$ values), if the deuteron scattering angle $\Theta_d^{lab}$ is
close to 
$0^{\circ}$. A possible albeit controversial explanation has been given in 
Ref. \cite{col} by associating this central maximum with $\eta$ and
$\pi\pi\pi$ production.
In our $M_{\pi\pi}$ data, which cover all angles
$\Theta_d^{lab} \ge 3^{\circ}$, we observe no apparent high-mass
enhancement.

The $\pi\pi$ low-mass enhancements observed in the exclusive data for the 
$\pi^0\pi^0$ channels are much larger than predicted in previous
$\Delta\Delta$ calculations \cite{ris,barn,anj,gar,mos}. As an example we show
in Figs. 2 - 4 (dashed lines) calculations in the model ansatz of
Ref. \cite{ris}, where we additionally included the pion angular 
distribution in $\Delta$ decay and  the Fermi smearing of the nucleons bound
in the final nucleus. We have chosen this model ansatz, since in addition to 
its simplicity it gives the smallest high-mass enhancement, i.e. is closest to
our observation in this respect.

\begin{figure}
\begin{center}
\includegraphics[width=0.45\textwidth]{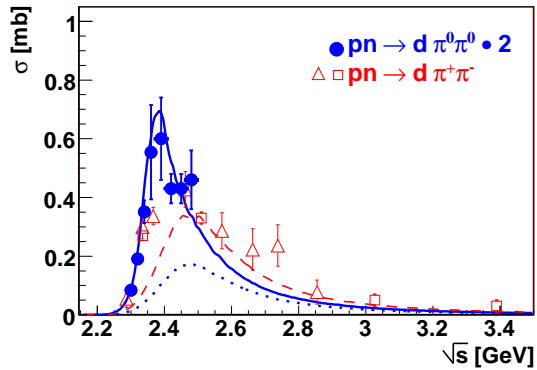}
\end{center}
\caption{Energy dependence of the total cross section for the $pn\rightarrow
  d\pi^+\pi^-$ reaction from threshold ($\sqrt{s}$ = 2.15 GeV) up to
  $\sqrt{s}$ = 3.5 GeV. Experimental data are from Refs. \cite{abd} (open
  squares) and \cite{bar} (open triangles). The results of this work for
  the $\pi^0\pi^0$ channel --- scaled by the isospin factor of 
   two --- are given by the full circles. Dashed and dotted lines
   represent the cross sections for $\pi^+\pi^-$ and $\pi^0\pi^0$
   channels, respectively, as expected from the isovector $\pi^+\pi^0$ data
   by isospin relations (see text). The solid curve includes a {\it s}-channel
   resonance in the 
   $\Delta\Delta$ system adjusted to describe the ABC effect in the
   $\pi^0\pi^0$ channel.}  
\end{figure}

Since on the one hand the available $\Delta\Delta$ calculations 
obviously fail, but on the other hand the data clearly show the $\Delta\Delta$
excitation in their $M_{N\pi}$ spectra, a profound physics piece appears to be 
missing in the interpretation. As we demonstrate in the following such a
missing piece is found by   
inspection of the energy dependence of the total cross section 
displayed in Fig. 4. Shown are the results of this work (solid circles) for
$pn \to d\pi^0\pi^0$ --- multiplied by the isospin factor of
2 --- in comparison with those of Refs. \cite{abd,bar} for the $pn \to
d\pi^+\pi^-$ reaction (open symbols). The latter reaction is composed of $I$ =
0 and 1 
contributions. The isovector part can be directly derived from the known $pp
\to d\pi^+\pi^0$ cross sections by use of isospin relations \cite{bys}. Since
the $pp \to d\pi^+\pi^0$ reaction is very well described \cite{FK} by
conventional $t$-channel calculations for the formation of a
$\Delta\Delta$ system in the intermediate state, also the isoscalar
part of this conventional process can be derived by applying isospin relations
to the intermediate $\Delta\Delta$ system. The thus obtained $I=0$ part is
shown in Fig. 4 by the dotted line, whereas the non-interfering sum of
isoscalar and isovector parts is given by the dashed line. Indeed, the
$\pi^+\pi^-$ data are in reasonable agreement with this sum for $\sqrt{s} \ge$
2.5 GeV. At lower energies, however, the measured values are much larger than
expected from the conventional $\Delta\Delta$ process. 


The so far unexplained structure observed for  $\sqrt{s} <$ 2.5 GeV in our
data is much larger and narrower 
than expected from the conventional $\Delta\Delta$ excitation by $t$-channel
meson exchange \cite{ris,barn,mos}. 
The cross section maximum is shifted substantially below the nominal 
$\Delta\Delta$ threshold being, however, still 200 MeV above the $d\pi\pi$
threshold.    


Since isospin is very unlikely to be broken on such a large scale, a mechanism
different from the t-channel $\Delta\Delta$ excitation must be the reason for
this resonance-like structure in the isoscalar sector. 
These considerations led us to the concept of a {\it s}-channel resonance, 
which couples to isoscalar $pn$ and $\Delta\Delta$ configurations. We note
that such a kind of dibaryon resonance has been predicted by a number of
theoretical investigations \cite{ping,barnes,kam,oka,kuk,mot}. 
%

For the description of the observables by such a resonance we use the
following Breit-Wigner ansatz for the resonance amplitude

$A_R \sim F(q_{\Delta\Delta})* D_R * D_{\Delta_1} *
D_{\Delta_2}$~~~~~~~~~~~~~~~~~~~~(1) 

with mass and width of the {\it s}-channel resonance being $m_R \approx$ 2.36
GeV/c$^2$ and $\Gamma_R \approx$ 80 MeV and   
where $D_R$ and $D_{\Delta}$ stand for {\it s}-channel resonance and  $\Delta$
propagators, respectively. 
The  form
factor $F(q_{\Delta\Delta})$ of the $\Delta\Delta$ vertex, which is chosen to
be of monopole type, depends on the relative momentum $q_{\Delta\Delta}$
between the two $\Delta$s. Since $q_{\Delta\Delta}$ = $q_{\pi\pi}$, when
neglecting the Fermi motion of the nucleons, this form factor is reflected
directly in the $M_{\pi\pi}$ spectra and 
causes there the ABC effect by suppression of the high-mass region. Fitting
the cutoff parameter $\Lambda_{\Delta\Delta}$ of this monopole formfactor to
the data in the  $M_{\pi\pi}$ spectra  results in
$\Lambda_{\Delta\Delta} \approx$ 0.2 GeV/c which corresponds to the
size of a deuteron-like object. 

While other theoretical explanations cannot be ruled out at this time, with
such a simple ansatz we obtain a surprisingly good description of the 
data (MC Dalitz 
plot in Fig.1 and solid lines in Figs. 2 - 4) both in their energy 
dependence and in their differential behavior.
 In previous attempts \cite{bash, menu07} of data 
interpretations we had to impose a strong attraction or even a boundstate
condition between the two $\Delta$s,
in order to obtain a satisfactory description of data on the
ABC effect. Having learned now from the behavior of the total cross
section, that possibly a s-channel resonance could be the cause, these
impositions would find a natural explanation.

We finally note that after the move of the WASA detector to COSY new data have
been taken there very recently on this issue with two orders of magnitude
higher statistics. It is expected that these will be able to finally solve the
long-standing ABC puzzle. 


We acknowledge valuable discussions with L. Alvarez-Ruso, V. Anisovich,
D. Bugg, L. Dakhno, C. Hanhart, M. Kaskulov, V. Kukulin, E. Oset,
I. Strakovsky, 
F. Wang, W. Weise  and C. Wilkin on this issue. 
This work has been supported by BMBF
(06TU201, 06TU261), COSY-FFE, 
DFG (Europ. Graduiertenkolleg 683)
and the Swedish Research Council. 


\begin{thebibliography}{9}
\bibitem{abc} N. E. Booth, A. Abashian, K. M. Crowe, Phys. Rev. Lett.
  {\bf 7}, 35 (1961) ; {\bf 6}, 258 (1960); Phys. Rev. {\bf C132}, 2296ff
  (1963)  
\bibitem{hom} R. J. Homer {\it et al.}, Phys. Rev. Lett.{\bf 9}, 72 (1964) 
\bibitem{hal} J. H.Hall  {\it et al.}, Nucl. Phys. {\bf B12},573 (1969)
\bibitem{bar} I. Bar-Nir {\it et al.}, Nucl. Phys. {\bf B54},17 (1973)
\bibitem{ban} J. Banaigs {\it et al.}, Nucl. Phys. {\bf B67}, 1 (1973)
\bibitem{ban1} J. Banaigs {\it et al.}, Nucl. Phys. {\bf B105}, 52 (1976)
\bibitem{plo} F.Plouin  {\it et al.}, Nucl. Phys. {\bf A302},413 (1978)
\bibitem{abd} A. Abdivaliev  {\it et al.}, Sov. J. Nucl. Phys.{\bf 29}, 796
  (1979) 
\bibitem{col}  F. Plouin, P. Fleury, C. Wilkin, Phys. Rev. Lett. {\bf 65}, 690
  (1990)
\bibitem{wur} R. Wurzinger  {\it et al.}, Phys. Lett. {\bf B445}, 423 (1999)
\bibitem{cod} for a review see A. Codino and F. Plouin, LNS/Ph/94-06  
\bibitem{ris} T. Risser and M. D. Shuster, {\it Phys. Lett.} {\bf 43B}, 68
  (1973) 
\bibitem{barn} I. Bar-Nir, T. Risser, M. D. Shuster, Nucl. Phys. {\bf B87}, 109
  (1975)  
\bibitem{anj} J. C. Anjos, D. Levy, A. Santoro, Nucl. Phys. {\bf B67}, 37
  (1973) 
\bibitem{gar} see, e.g., A. Gardestig, G. F\"aldt, C. Wilkin, Phys. Rev.
  {\bf C59}, 2608 (1999) and Phys. Lett. {\bf B421}, 41 (1998)
\bibitem{mos}  C. A. Mosbacher, F. Osterfeld, nucl-th/9903064 
\bibitem{alv} L. Alvarez-Ruso, {\em Phys. Lett.} {\bf B452}, 207 (1999); PhD
  thesis, Univ. Valencia 1999  
\bibitem{barg} Chr. Bargholtz {\it et al.}, Nucl. Instr. Meth {\bf A594}, 339
  (2008)
\bibitem{bash} M. Bashkanov {\it et al.}, Phys. Lett. {\bf B637} 223 (2006) 
\bibitem{MB} M. Bashkanov, PhD thesis, Univ. T\"ubingen  2006,
\bibitem{sam} S. Keleta, PhD thesis, Uppsala Univ. 2008 
\bibitem{bys} J.Bystricky  {\it et al.}, J. Physique {\bf 48}, 1901 (1987)
\bibitem{FK} F. Kren  {\it et al.}, Proc. MESON08, in press
\bibitem{ping} J. Ping  {\it et al.}, {\em Phys. Rev.} {\bf
    C65}~044003~(2002) and references therein; arXiv:0806.0458
\bibitem{barnes} T. Barnes {\it et al.}, {\em Phys. Rev.} {\bf
    C48}~539~(1993)
\bibitem{kam} T. Kamae and T. Fujita,  {\em Phys. Rev. Lett.} {\bf 38}, 471
  (1977) 
\bibitem{oka} M.~Oka, K.~Yazaki, {\em Prog.~Theor.~Phys.} {\bf 66}~572~(1981)  
\bibitem{kuk} V. I. Kukulin  {\it et al.}, {\em Nucl. Phys.} {\bf A689} 327c
  (2001) 
\bibitem{mot} R. D. Mota {\it et al.}, {\em Phys. Rev.} {\bf C65} 034006
  (2002)
\bibitem{menu07} M.Bashkanov {\it et al.}, eConf C070910, 129
  (2007) 

\end{thebibliography}
\end{document}